\documentstyle[preprint,eqsecnum,aps]{revtex}

\begin{document}
\draft
\title{A Model of the Universe including dark Energy}
\date{\today}
\author{Rolando Cardenas\thanks{rcardenas@mfc.uclv.edu.cu}, Tame Gonzalez\thanks{tame@mfc.uclv.edu.cu}, Osmel Martin\thanks{osmel@mfc.uclv.edu.cu}and Israel Quiros\thanks{israel@mfc.uclv.edu.cu}}
\address{Departamento de Fisica. Universidad Central de Las Villas. Santa Clara. CP: 54830 Villa Clara. Cuba}
\maketitle

\begin{abstract}
In this work we explore a model of the universe in which dark energy is modelled explicitely with both a dynamical quintessence field (with a double exponential self-interaction potential) and a cosmological constant. For a given region of the parameter space, our results confirm the possibility of a collapsing universe, which is necessary for an adequate definition of both perturbative quantum field and string theories. We have also reproduced the measurements of modulus distance from supernovae with good accuracy.  
\end{abstract}

\section{Introduction}

From 1998 to date several important discoveries in the astrophysical
sciences have being made, which have given rise to the so called New
Cosmology \cite{turner1,turner2}. Amongst its more important facts we
may cite: 

- Flat, critical density accelerating universe

- Early period of rapid expansion (inflation)

- Density inhomogeneities produced from quantum fluctuations during
inflation

- Composition: 2/3 dark energy; 1/3 dark matter; 1/200 brigh stars

- Matter content: $(29\pm 4)\%$ nonbarionic dark matter; $(4\pm 1)\%$
			  baryons, $(0.1-5)\%$ neutrinos

- $T_0=2.275\pm 0.001 K$

- $t_0=14\pm 1 Gyr$

- $H_0=72\pm 7 km.s^{-1} Mpc^{-1}$ 

 The leading candidates to be identified with dark energy involve
fundamental physics and include a cosmological constant (vacuum energy), a
rolling scalar field (quintessence), and a network of light, frustrated
topological defects \cite{mst}.
 
Though probably there is a long way to go, the increasing amount of astrophysical data is providing tools (for instance, orthogonal tests) to break the known degeneracy of sets of cosmological parameters, increasing the accuracy of their determination. In some works, the use of several different data sets seems to favour a cosmological constant behaviour for today's dark energy, though the results could be plagued by theoretical assumptions \cite{bm}.

On the other hand, an eternally accelerating universe seems to be at odds
with string theory, because of the impossibility of formulating the
S-matrix. In a deSitter space the presence of an event horizon, signifying
causally disconnected regions of space, implies the absence of asymptotic
particle states which are needed to define transition amplitudes \cite
{banks,cline}. Also in the context of perturbative quantum field theory a horizon problem appears, called superexpansionary divergence \cite{ssyy}.

Exponential potentials are natural in theories with extra compact dimensions, such as Kaluza-Klein supergravity and superstring models \cite{halliwell,sw}. Exponential quintessence has been studied extensively in the literature and often discarded from limits on big-bang nucleosynthesis or on fine tuning arguments. This is the case when a fixed point solution for the quintessence evolution is reached early on in the universe, but it is possible that we have not yet reached this fixed point. For instance, in \cite{lr} it is shown that still there are reasonable regions of the parameter space of the simple exponential potential model for quintessence that are allowed by some observational constraints, and that the required level of fine tuning is not too stringent. Furthermore, several authors have recently pointed out that the degree of
fine tuning needed in these scenarios is no more than in others usually
accepted \cite{cline,kl,rubano}.

Due to the above there is a renewed interest in exponential quintessence, largely 
because in several scenarios exponential potentials can reproduce the
present acceleration and predict future deceleration, avoiding the horizon problem that appears in the context of both perturbative quantum field and string theories.

The cosmological constant can be incorporated into the quintessence
potential as a constant which shifts the potential value, especially, the
value of the minimum of the potential, where the quintessence field rolls
towards. Conversely, the height of the minimum of the potential can also be
regarded as a part of the cosmological constant. Usually, for separating
them, the possible nonzero height of the minimum of the potential is
incorporated into the cosmological constant and then set to be zero. The
cosmological constant can be provided by various kinds of matter, such as
the vacuum energy of quantum fields and the potential energy of classical
fields and may also be originated in the intrinsic geometry. So far there is
no sufficient reason to set the cosmological constant (or the height of the
minimum of the quintessence potential) to be zero \cite{hwang}. In
particular, some mechanisms to generate a negative cosmological constant
have been pointed out, in the context of spontaneous symmetry breaking \cite{ss,gh}, although the combined data of the 2dFGRS (two-degree Field Galaxy Redshift Survey) and CMB anisotropies suggest a positive cosmological constant \cite{efstathiou}.

The goal of this paper is to explore a model of the universe in which the dark energy component is accounted for by both a quintessence field and a negative cosmological constant. The net result of this combination would be a possitive "effective" cosmological constant. In a former paper \cite{cglmq} we explored this scenario considering a single exponential potential for the quintessence field. However, several authors have shown that the double exponential potential tends to give a better agreement with experimental data \cite{rubano,cc}. The quintessence field accounts for the present stage of accelerated expansion of the universe. Meanwhile, the inclusion of a negative cosmological constant warrants that the present stage of accelerated expansion will be, eventually, followed by a period of collapse into a final cosmological singularity (AdS universe). 

\section{The Model}

We consider a model consisting of a three-component cosmological
fluid: matter, scalar field (quintessence with an exponential potential) and
cosmological constant. ''Matter'' means barionic + cold dark matter, with no
pressure, and the scalar field is minimally coupled and noninteracting with
matter. This model cannot be used from the very beginning of the universe,
but only since decoupling of radiation and dust. Thus, we don't take into
account inflation, creation of matter, nucleosynthesis, etc. 

The action of the model under consideration is given by

\begin{equation}
S=\int d^4 x\;\sqrt{-g}\{\frac{c^2}{16\pi G}(R-2\Lambda)+{\cal L}_\phi+{\cal %
L}_{m}\},
\end{equation}
where $\Lambda$ is the cosmological constant, ${\cal L}_{m}$ is the
Lagrangian for the matter degrees of freedom and the Lagrangian for the
quintessense field is given by

\begin{equation}
{\cal L}_{\phi}=-\frac{1}{2}\phi_{,n} \phi^{,n}-V(\phi).
\end{equation}

\bigskip Now we
apply the same technique of adimensional variables we proposed in \cite{cmq} to
determine the integration constants without additional assumptions.
We use the dimensionless time variable $\tau =H_{0}t$, where $t$ is
the cosmological time and $H_{0}$ is the present value of the Hubble
parameter. In this case $a(\tau )=\frac{a(t)}{a(0)}$ is the scale factor. Then we have that, at present $(\tau =0)$

\begin{eqnarray}
a(0)&=&1,  \nonumber \\
\dot{a}(0)&=&1,  \nonumber \\
H(0)&=&1,
\end{eqnarray}

\bigskip Considering a homogeneous and isotropic universe and using
the experimental fact of a spatially flat universe \cite{bernardis}, 
the field equations derivable from (2.1) are

\begin{equation}
(\frac{\dot{a}}{a})^2=\frac{2}{9}\sigma^2 \{ \frac{\bar D}{a^3}+\frac{1}{2} 
\dot{\phi}^2 + \bar{V}(\phi)+\frac{3}{2}\frac{\bar{\Lambda}}{\sigma^2}\},
\end{equation}

\begin{equation}
2\frac{\ddot{a}}{a}+(\frac{\dot{a}}{a})^2=-\frac{2}{3}\sigma^2 \{ \frac{1}{2}%
\dot{\phi}^2-\bar{V}(\phi)-\frac{3}{2}\frac{\bar{\Lambda}}{\sigma^2}\},
\end{equation}
and

\begin{equation}
\ddot{\phi}+3\frac{\dot{a}}{a}\dot{\phi}+\bar{V}^{\prime}(\phi)=0,
\end{equation}
where the dot means derivative in respet to $\tau$ and,

\begin{eqnarray}
\bar{V}(\phi)&=&\bar{A}^2 e^{\sigma\phi} + \bar{B}^2 e^{-\sigma\phi},
\end{eqnarray}

\begin{equation}
\bar{X}^{2} = \frac{X^{2}}{H_{0}^{2}},  
\end{equation}
except for
\begin{equation}
\bar{D} = \frac{D}{a_{0}^{3}H_{0}^{2}}=\frac{\rho _{m_{0}}}{H_{0}^{2}},
\end{equation}
with $\rho _{m_{0}}$ - the present density of matter, $\sigma ^{2}=\frac{12\pi G%
}{c^{2}}$, and $A^{2}$ and $B^{2}$ - generic constants.

Applying the Noether Symmetry Aproach \cite{rmrs,r,crrs,rr}, it can be shown that the new variables we should introduce to simplify the
field equations are the same used in \cite{rubano}: 

\begin{equation}
a^3=\frac{u^2-v^2}{4},
\end{equation}
and

\begin{equation}
\phi =\frac{1}{\sigma }\ln [\frac{B(u+v)}{A(u-v)}].
\end{equation}

\bigskip Now the field equations take the form:
\begin{equation}
\ddot{u}=(\bar{A}\bar{B}\sigma ^{2}-\sigma ^{2}\bar{V}_{0}),
\end{equation}
and
\begin{equation}
\ddot{v}=-(\bar{A}\bar{B}\sigma ^{2}+\sigma ^{2}\bar{V}_{0})v.
\end{equation}
\bigskip
where 
\begin{equation}
\bar{V}_{0}=-\frac{3}{2}\frac{\bar{\Lambda}}{\sigma^2}
\end{equation}
In what follows, we introduce the adimensional densities $\Omega_i$ (in units of the critical density of the universe). Subscript $i$ = $m$, $Q$ or $\Lambda$ for matter, quintessence field and cosmlogical constant, respectively. Subscript $0$ represents present values (except for $\bar{V}_{0}$). We also introduce $q$, the deceleration parameter of the universe.
Now, for definiteness, we assume the following relation between the constants:
\begin{equation}
\bar{A}^{2}=n\bar{B}^{2}=m\bar{V}_{0},
\end{equation}
where $n$ and $m$ are real parameters.
\bigskip

Then, the solutions of the equations (2.12) and (2.13) are
\begin{equation}
u(\tau )=b\exp [-\sqrt{-\frac{9}{2}\Omega _{\Lambda }(m\sqrt{\frac{1}{n}}-1)%
}t]+c\exp [\sqrt{-\frac{9}{2}\Omega _{\Lambda }(m\sqrt{\frac{1}{n}}-1)}t],
\end{equation}

\begin{equation}
v(\tau )=d\sin [\sqrt{-\frac{9}{2}\Omega _{\Lambda }(m\sqrt{\frac{1}{n}}+1)}%
t]+e\cos [\sqrt{-\frac{9}{2}\Omega _{\Lambda }(m\sqrt{\frac{1}{n}}+1)}t],
\end{equation}
\bigskip

In finding the integration constants we use the equations (2.3) and field equations evaluated at $\tau =0$. Finally, using $\Omega_{m_0}+\Omega_{Q_0}+\Omega_{%
\Lambda}=1 $, the integration constants are given through
\bigskip

\begin{equation}
e=-\sqrt{\frac{1-\Omega _{m_{0}}-4\Omega _{\Lambda }+\frac{1-2q_{0}}{3}}{%
-2\Omega _{\Lambda }m\sqrt{\frac{1}{n}}}-2},
\end{equation}
\begin{equation}
c=\frac{-e\sqrt{6(2-3\Omega _{m_{0}}+2q_{0})}-6\sqrt{e^{2}+4}}{8\sqrt{-%
\frac{9}{2}(m\sqrt{\frac{1}{n}}-1)\Omega _{\Lambda }}}+\frac{\sqrt{e^{2}+4}}{%
2},
\end{equation}
\begin{equation}
b=\sqrt{e^{2}+4}-c,
\end{equation}
\begin{equation}
d=\frac{-6+(b+c)(b-c)\sqrt{-\frac{9}{2}(m\sqrt{\frac{1}{n}}-1)\Omega
_{\Lambda }}}{e\sqrt{-\frac{9}{2}(m\sqrt{\frac{1}{n}}+1)\Omega _{\Lambda }}}.
\end{equation}
\bigskip

\section{Analysis of results}

Our constants (and, consequently, the solutions) depend on five
physical parameters: $\Omega _{m_{0}}$, $\Omega _{\Lambda }$, $q_{0}$ and on 
the positive real numbers $m$ and $n$.

Concerning the values of these parameters, since $\sqrt{6(2+2q_{0}-3\Omega _{m_{0}})}$ should be real (see equation
(2.19)) then, the following constrain on the present value of the
deceleration parameter follows

\begin{equation}
q_0\geq 1.5\Omega_{m_0}-1.
\end{equation}
 
It was determined that the only relevant cosmological magnitude that has a sensible dependence on parameters $m$ and $n$ is the state parameter $\omega$. We used $\Omega_{m_{0}}=0.3$. Though we made calculations for several values of $\ \Omega _{\Lambda }$ in the range -0.01 a -0.30, for simplicity we present results for -0.15, bearing in mind that they change little for other values.

Figure 1 shows the evolution of the scale factor for $\Omega _{\Lambda
}=-0.15$. It can be shown that, in other to have a collapsing universe, the following condition should be fulfilled (otherwise there is eternal expansion):
\begin{equation} 
m<\frac{1}{\sqrt{n}}
\end{equation}
We also saw that with the decrease (modular increase)
of $\ \Omega _{\Lambda }$, the time of collapse diminishes.

We got the dependence of the state parameter and of
the deceleration parameter with the redshift and them we selected the values of $m$, $n$ and $q_{0}$. For this purpose we considered the results
of Turner and Riess\cite{tr}.

Figure 2 shows the behaviour of the deceleration parameter as function of the
redshift z for the
same values of the parameters. This figure shows an early stage of deceleration and a current epoch of acceleration. A transition from the decelerated phase
to the acelerated one is seen approximately for z =0.5. We appreciate an increase of
the deceleration parameter upon increasing the value of z. This points at a past epoch in the evolution when gravity of the dark energy was attractive. As follows from figure 1, aceleration is not eternal: in the future $q>0$ again,
which gives rise to the collapse.

Figure 3 shows the evolution of the state parameter of the effective quintessence field $\omega
_{\phi }$. It's noticeable that the effective quintessence field has state
parameter $\omega_{\phi }$ near $-0.8$ today, which means that its behaviour is close to the pure cosmological constant, as a vacuum fluid. This is in agreement with the results of \cite{cc} when only the first Doppler peak is considered.
If we are to explain the
past and future deceleration obtained
in our model, it's important to look at the dynamical quintessence field. We
see that in the recent past $\omega_{\phi }>0$, which implies that
quintessence field behaved (or simply was) like ordinary atractive matter,
giving rise to the logical deceleration. In the future this will happen
again ($\omega_{\phi }>0)$ , with the consequent deceleration.

The present values
of the physical parameters ($\Omega _{m_{0}}=0.3$, $\Omega _{\Lambda }=-0.15$,
and $q_{0}=-0.34$, were chosen after a detailed analisys of the behaviour of these parameters shown in figs. 2 and 3.

Now we proceed to analyze how our solution reproduces some experimental results.
With this purpose, in Fig. 4 we plot the distance modulus $\delta (z)$ vs
redshift z, calculated by us and the one obtained with the usual model with
a constant $\Lambda $ term. The relative deviations are of about 0.3$\%$, which improves the 0.5$\%$ obtained by us in \cite{cglmq} for the single exponential potential.

\section{Conclusions}
In a recent paper \cite{hwang} it is pointed out that the ultimate fate of our
universe is much more sensitive to the presence of the cosmological constant
than any other matter content. In particular, the universe with a negative
cosmological constant will always collapse eventually, even though the
cosmological constant may be nearly zero and undetectable at all at the
present time. It is also interesting that Sen and Sethi, using an ansatz for the scale factor that produces future deceleration, obtain from the field equations that the quintessence potential should be a double exponential plus a constant \cite{sese}.Our results also support the very general
assertions of \cite{hwang}, we have shown that for a determined region of the parameter space, the universe collapses. This also favours the formulation of string and quantum field theories, as explained in the introduction. 
The experimental measurements of modulus distance from the supernovae are adequately reproduced within an accuracy of 0.3$\%$. 
So far, we have investigated one of the several possible branches of the solution, leaving for the future the investigation of the others. We have also reserved for future work the careful examination of this universe near its beginning (i.e., just after the decoupling of matter and radiation), and the proper constrain of the parameters of this model using several data sets.

We acknowledge Claudio Rubano, Mauro Sereno and Paolo Scudellaro, from Universita di Napoli "Federico II" , Italy, for useful comments and discussions and Andro Gonzales for help in the computations. We also acknowledge Alessandro Melchiorri, from Oxford University, for drawing our attention to reference \cite{bm}.

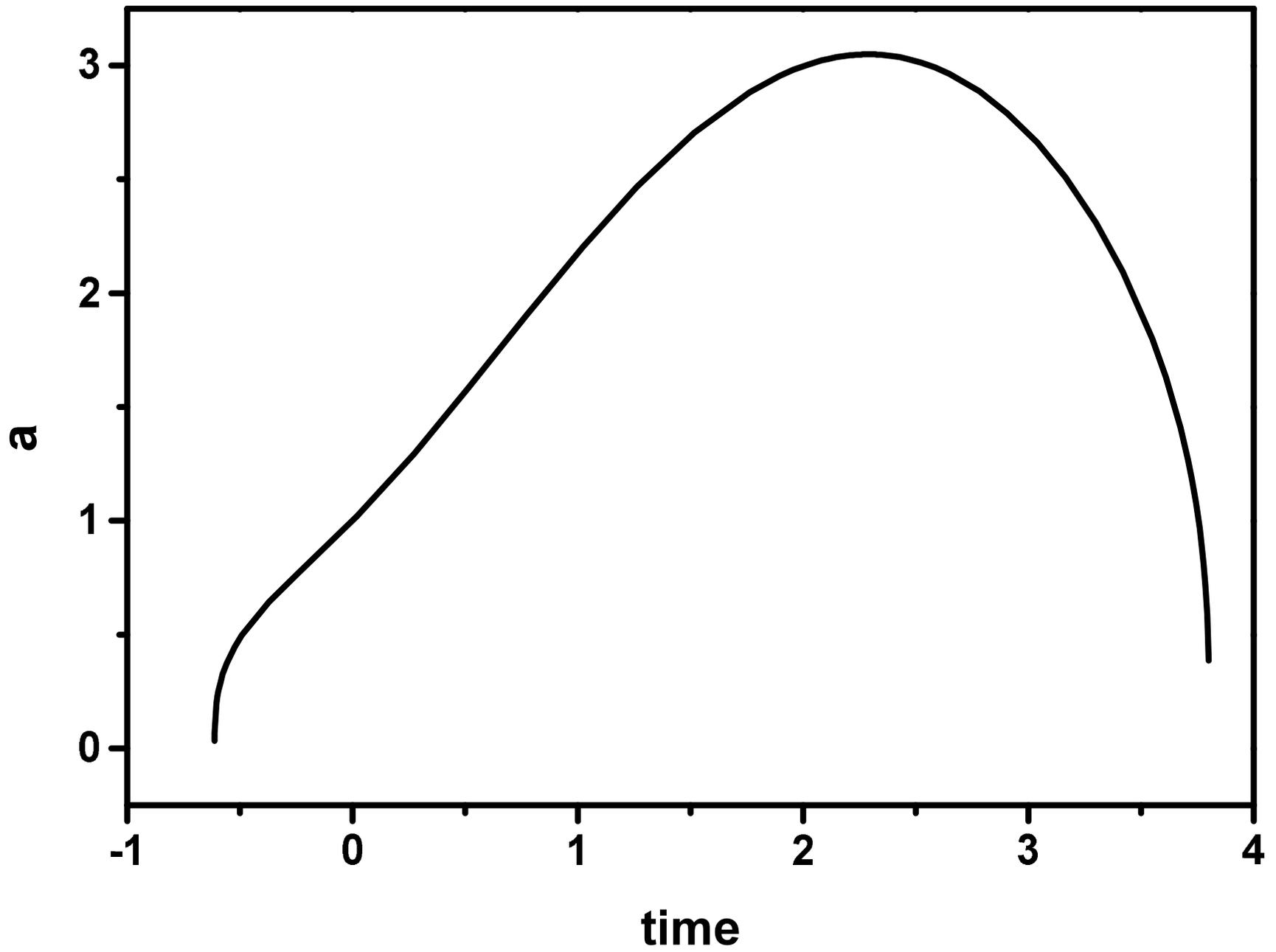
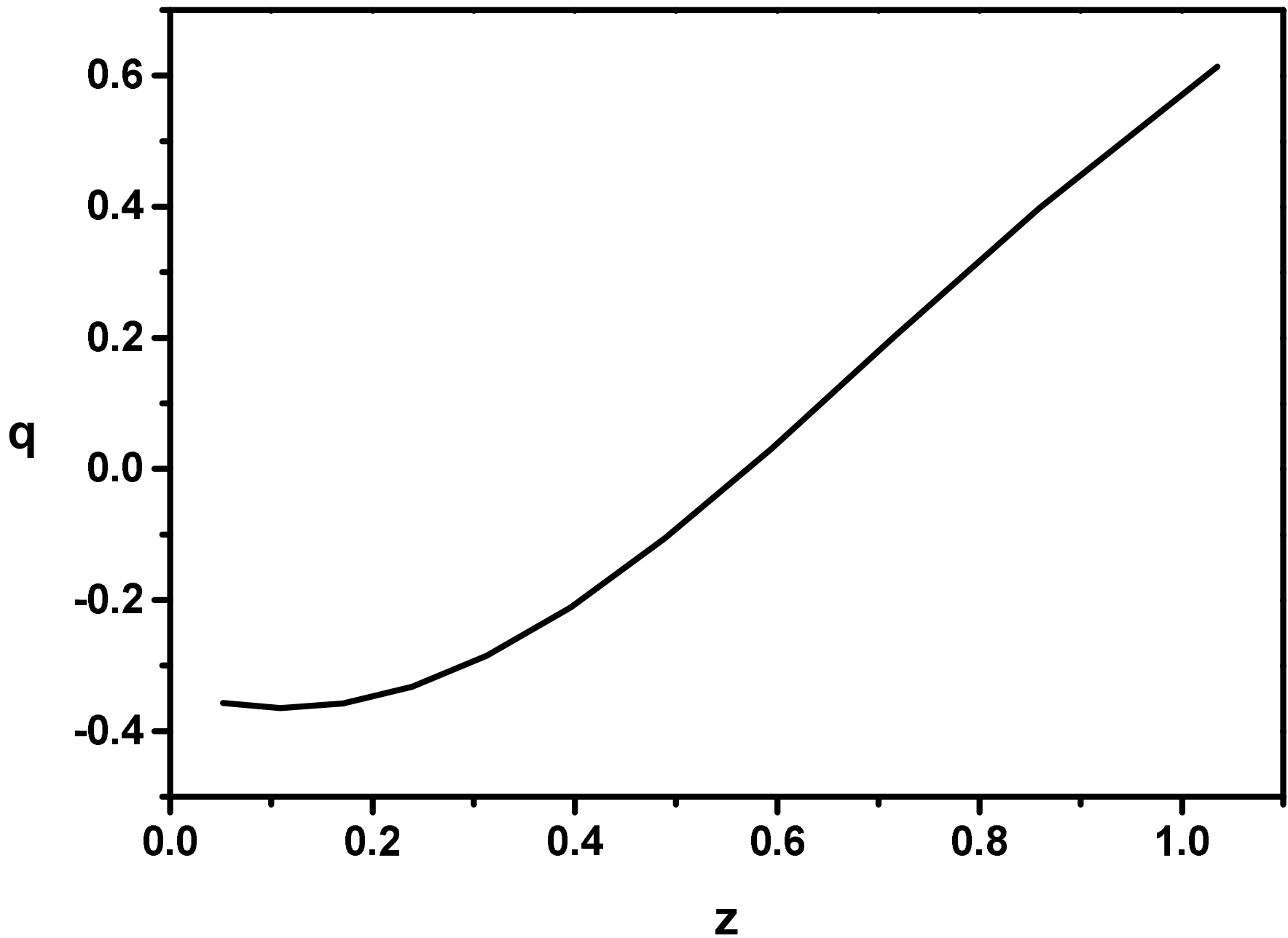
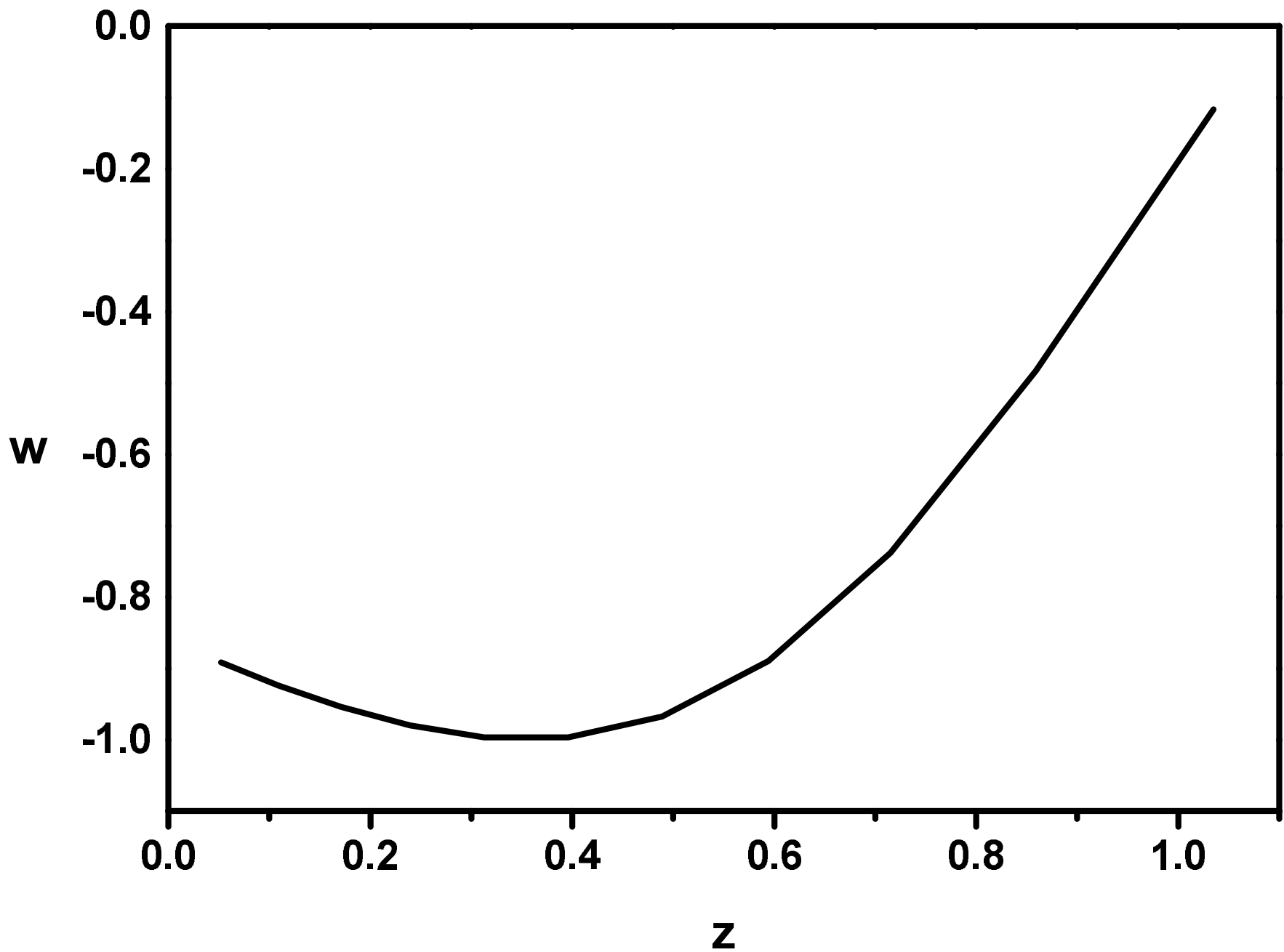
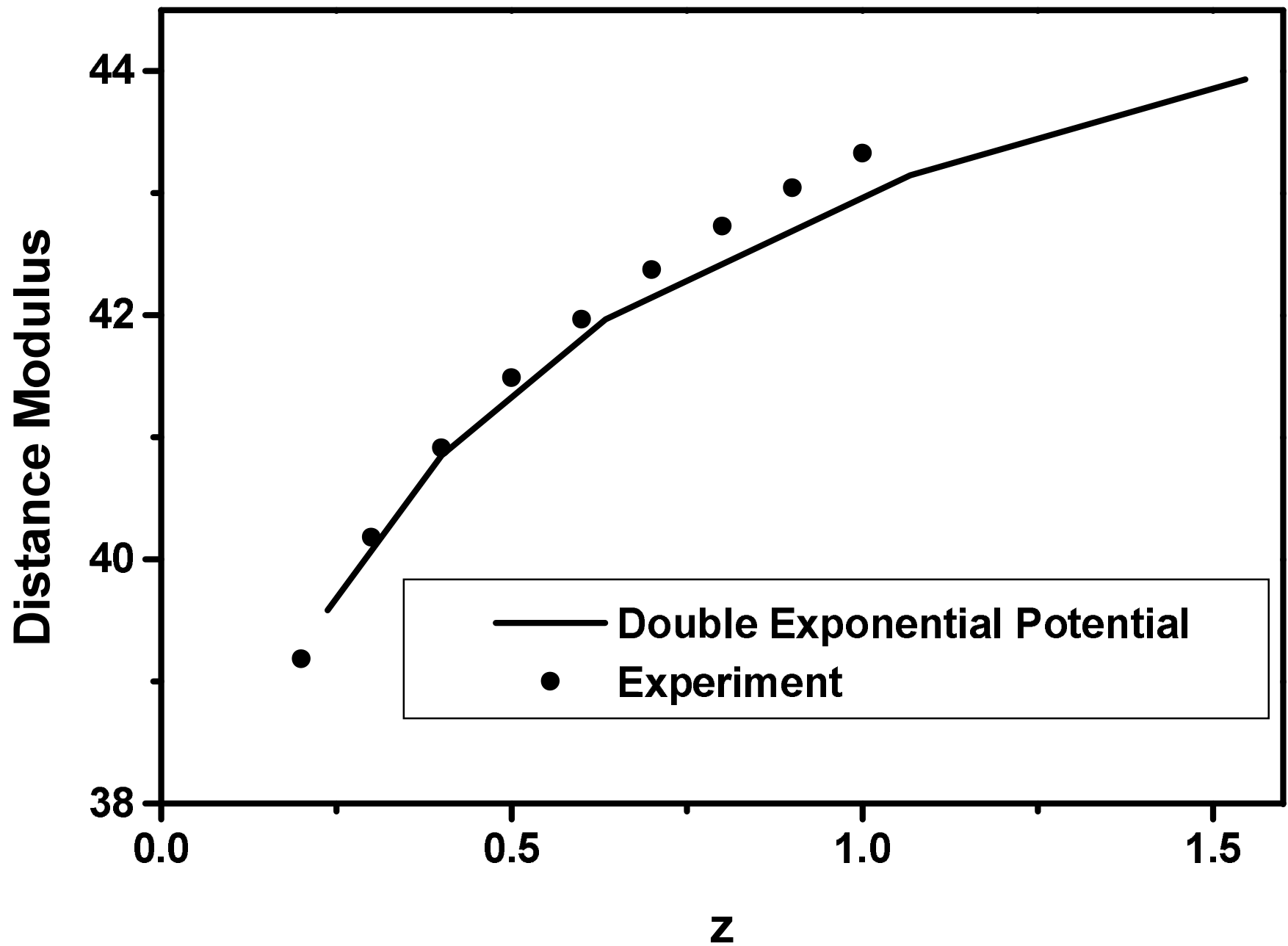
\end{document}